\documentstyle[12pt,hpatex,epsf]{article}

  \newcommand{\de}{\delta}

  \newcommand{\ga}{\gamma}

  \newcommand{\Om}{\Omega}
  \newcommand{\om}{\omega}
  \newcommand{\si}{\sigma}
  
  \newcommand{\th}{\theta}

  \newcommand{\lap}{\triangle}
  
  \newcommand{\bm}[1]{\mbox{\boldmath $#1$}}
  \newcommand{\be}{\begin{equation}}
  \newcommand{\ee}{\end{equation}}

  \newcommand{\bea}{\begin{eqnarray}}
  \newcommand{\eea}{\end{eqnarray}}
  \newcommand{\bean}{\begin{eqnarray*}}
  \newcommand{\eean}{\end{eqnarray*}}
  \newcommand{\dd}{\partial}
  
  \newcommand{\cd}{\cdot}

\begin{document}          
   \title{ Anisotropies in the Cosmic Microwave 
  Background: Theoretical Foundations}
  \authors{Ruth Durrer}
 \address{D\'epartement de Physique Th\'eorique, Universit\'e de
  Gene\`eve,\\ 24, quai E. Ansermet, CH-1211 Gen\`eve 4}
\abstract{
 The analysis of anisotropies in the cosmic microwave background (CMB) has
   become an extremely valuable tool for cosmology. We  even have
   hopes  that planned CMB anisotropy experiments may revolutionize
   cosmology. Together with determinations of the CMB spectrum, they
   represent the first cosmological precision measurements. This is 
   illustrated  in  the talk by Anthony 
   Lasenby. The value of CMB anisotropies lies to a big part in
   the  simplicity of the theoretical analysis. Fluctuations in the CMB
   can be determined almost fully within linear cosmological
   perturbations theory and are not severely influenced by complicated
   nonlinear physics. 

  In this contribution   the different physical processes causing
  or influencing anisotropies in the CMB are discussed. 
  The geometry perturbations at
  and after last scattering, the acoustic oscillations in the 
  baryon--photon--plasma prior to recombination, and the diffusion 
  damping during the process of recombination. 

  The perturbations due to the fluctuating gravitational field, the
  so called Sachs--Wolfe contribution, is described in a very general
  form using the Weyl tensor of the perturbed geometry. 
}

  \section{Introduction}

  The formation of cosmological structure in the universe, 
   inhomogeneities in the matter distribution like quasars at redshifts 
  up to $z\sim 5$, galaxies, clusters, 
  super clusters, voids and walls,  is an outstanding basically 
   unsolved problem within the standard model of cosmology. 
  We assume, that the observed inhomogeneities formed from small initial
   fluctuations by gravitational clustering.

  At first sight it seems obvious that small density enhancements can grow
  sufficiently rapidly by gravitational instability. But global 
  expansion of the universe and
  radiation pressure counteract gravity, so that, e.g., in the case of
  a radiation dominated, expanding universe no density inhomogeneities
  can grow significantly. Even in a universe dominated by 
  pressure-less matter, cosmic dust,
  growth of density perturbations is strongly reduced by the expansion
  of the universe.

  Furthermore, we know that the universe was extremely homogeneous
  and isotropic at early times. This follows from the isotropy of
  the 3K Cosmic Microwave Background (CMB), which represents a relic of
  the  plasma of baryons, electrons and radiation at times before protons 
  and electrons combined to neutral hydrogen. After a long series of 
  upper bounds,
  measurements with  the DMR instrument aboard the  COsmic Background 
  Explorer satellite  (COBE)
  have finally established anisotropies in this radiation \cite{Sm} at 
  the level of 
  \[\left<{(T(\bm{n})-T(\bm{n}'))^2\over T^2}\right>_
	{({\bf\small n\cd n'}=cos\th)} \sim 10^{-10} 
	  ~~~\mbox{ on  angular scales}~
		   7^o\le \theta\le 90^o ~.\]

  Such an angle independent spectrum of fluctuations on large angular scales is called
  Harrison Zel'dovich spectrum \cite{HZ}. It is defined by yielding 
  constant  mass fluctuations on horizon scales at all time, i.e.,
   if $l_H(t)$
  denotes the expansion scale  at time $t$,
  \[ \langle(\Delta M/M)^2(\lambda=l_H)\rangle = \mbox{ const. , ~
  independent of time.} \]
  The COBE result, the observed spectrum and amplitude of fluctuations, 
  strongly support the gravitational instability picture.

  Presently, there exist two main classes of models which predict a
  Harrison--Zel'dovich spectrum 
  of primordial fluctuations: In the first class, quantum fluctuations 
  expand to super Hubble scales during a period of inflationary expansion 
  in the very early universe and `freeze in' as classical fluctuations in 
  energy density and geometry \cite{Slava} ({\sl see also the
  contribution by V. Mukhanov}).
  In the second class, a phase transition in the early universe, at a 
  temperature of about $10^{16}$GeV leads to 
  topological defects which induce perturbations in the geometry and in the 
  matter content of the universe \cite{Kibble}.
  Both classes of models are in basic agreement with the COBE findings,
  but differ in their prediction of anisotropies on smaller angular
  scales. 

  On smaller angular scales  the observational situation is 
  at present somewhat confusing and contradictory \cite{mor,La}, but many 
  anisotropies have been measured with a maximum  of about 
  $ \Delta T/T \approx (3\pm 2)\times 10^{-5} $ at angular scale 
  $\theta \approx (1\pm 0.5)^o$. There is justified hope, that the
  experiments planned and under way will improve this situation within
  the next few years (see contribution by A. Lasenby) In Fig.~1, the
  experimental situation as of spring '96 is presented.

\begin{figure}[htb]
\centering
\epsfysize=10.5cm
\epsffile{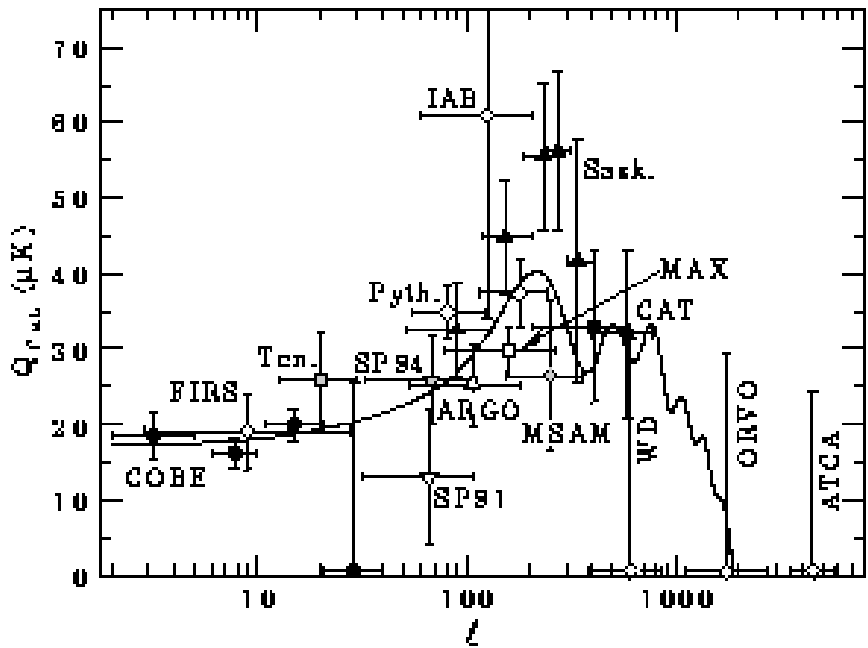}
\caption{\small The corresponding quadrupole amplitude $Q_{flat}$ is 
shown versus  the corresponding spherical harmonic index $\ell$. The 
amplitude $Q_{flat}(\ell)$ corresponds roughly to the temperature 
fluctuation on the angular scale $\th \sim \pi/\ell$. The solid line 
indicates the predictions from a standard cold dark matter model.
 (Figure taken from ref. [5]).}
\end{figure}

In this paper we outline a formal derivation of general formulas
which can be used to calculate the CMB anisotropies in a given
cosmological model. Since we have the chance to address a community of 
relativists, we make full use of the relativistic
formulation of the problem. In Section~2 we derive Liouville's
equation for massless particles in a perturbed Friedmann universe. In
Section~3 we discuss the effects of non-relativistic Compton
scattering prior to decoupling. This fixes the initial conditions for
the solution to the Liouville equation and leads to a simple 
approximation of the effect of collisional damping. In the next Section 
we illustrate our
results with a few simple examples. Finally, we summarize our
conclusions. 
\vspace{0.2cm}

{\bf Notation:} We denote conformal time by $t$. Greek indices run from 
0 to 3, Latin indices run from 1 to 3. The metric signature is chosen 
$(- + + +)$. The Friedmann metric is thus given by
$ds^2=a^2(t)(-dt^2+\ga_{ij}dx^idx^j)$, where $\ga$ denotes the metric
of a 3--space with constant curvature $K$. Three dimensional vectors are 
denoted by bold face symbols.\\
We set $\hbar=c=k_{Boltzmann}=1$ throughout.

\section{The Liouville equation for massless particles}
\subsection{Generalities}
Collision-less particles are described by their one particle 
distribution function which lives on the seven dimensional phase space 
\[ {\cal P}_m = \{ (x,p)\in T\!{\cal M}| g(x)(p,p)=-m^2\} ~. \]
Here $\cal M$ denotes the spacetime manifold and $T{\cal M}$ its 
tangent space. The fact that collision-less particles move on geodesics
translates to the Liouville equation for the one particle distribution
function, $f$. The Liouville equation reads \cite{Ste}
\begin{equation}  X_g(f)=0    \label{L}~.\end{equation}
In a tetrad basis $(e_\mu)_{\mu=0}^3$ of $\cal M$, the vector field
 $X_g$ on ${\cal P}_m$ is given 
by ({\sl see}, e.g., \cite{Ste})
\begin{equation} 
	X_g = (p^\mu e_\mu - \omega^i_{\:\mu}(p)p^\mu{\partial \over \partial p^i}) 
  ~, \label{liou}	\end{equation}
where $\omega^\nu_{\:\mu}$ are the connection 1--forms of $({\cal M},g)$ 
in the basis $e^\mu$, and we have chosen the basis  
\[(e_\mu)_{\mu=0}^3~~ \mbox{ and }~~~
({\partial\over \partial p^i})_{i=1}^3 ~~\mbox{  on }~~~~  T{\cal P}_m~, 
	~~~~~~  p=p^\mu e_\mu~. \]

We  now show that for massless particles and conformally related metrics,
\[g_{\mu\nu}= a^2\tilde{g}_{\mu\nu}~,\]
\begin{equation}
	(X_gf)(x,p)=0 ~~~\mbox{ is equivalent to }~~~
	(X_{\tilde{g}}f)(x,ap)=0  ~. 
	\label{conform}		\end{equation}
This is easily seen if we write $X_g$ in a coordinate basis:
\[ X_g = b^\mu\partial_\mu -\Gamma_{\alpha\beta}^ib^\alpha b^\beta{\partial\over \partial b^i}
	~,\]
with 
\[ \Gamma_{\alpha\beta}^i={1\over 2}g^{i\mu}(g_{\alpha\mu},_\beta + 
	g_{\beta\mu},_\alpha-g_{\alpha\beta},_\mu) ~. \]
The variables $b^\mu$ are the components of the momentum $p$ with respect to the
{\em coordinate} basis:
\[ p=p^\mu e_\mu = b^\mu\partial_\mu ~.\]
If $(e_\mu)$ is a tetrad with respect to $g$, then $\tilde{e}_\mu=ae_\mu$
is a tetrad basis for $\tilde{g}$. Therefore,  the coordinates of 
of $ap=ap^\mu\tilde{e}_\mu= a^2p^\mu e_\mu = a^2b^\mu\partial_\mu$, with 
respect to the basis $\partial_\mu$ on 
$({\cal M},\tilde{g})$ are  given by $a^2b^\mu$.
In the coordinate basis thus our statement Eq.~(\ref{conform}) follows, if we
can show that
\begin{equation}
	 (X_{\tilde{g}}f)(x^\mu,a^2b^i)=0 
	~~~\mbox{ iff }~~~  (X_gf)(x^\mu,b^i)=0 
	\label{conformco} \end{equation}
Setting $v=ap=v^\mu\tilde{e}_\mu = w^\mu\partial_\mu$, we have $v^\mu=ap^\mu$
and $w^\mu=a^2b^\mu$. Using $p^2=0$, we obtain the following relation 
for the Christoffel symbols of $g$ and $\tilde{g}$:
\[ \Gamma^i_{\alpha\beta}b^\alpha b^\beta=  \tilde{\Gamma}^i_{\alpha\beta}b^\alpha b^\beta
  +{2a,_\alpha\over a}b^\alpha b^i  ~.\]
For this step it is crucial that the particles are massless! For massive
particles the statement is of course not true.
Inserting this result into the Liouville equation we find
\begin{equation}
 a^2X_gf = w^\mu(\partial_\mu f|_b -2{a,_\mu\over a} b^i{\partial f\over \partial b^i})
  -\tilde{\Gamma}_{\alpha\beta}^iw^\alpha w^\beta {\partial f\over \partial w^i} ~,
	\label{lstar} \end{equation}
where $\partial_\mu f|_b$ denotes the derivative of $f$ w.r.t. $x^\mu$ at
constant $(b^i)$. Using
\[\partial_\mu f|_b = \partial_\mu f|_w + 2{a,_\mu\over a}b^i{\partial f\over \partial b^i} 
	~,\]
we see, that the  braces in Eq.~(\ref{lstar}) just correspond to 
$\partial_\mu f|_w$. Therefore,
\[a^2X_g f(x,p)=w^\mu\partial_\mu f|_w-\tilde{\Gamma}_{\alpha\beta}^iw^\alpha w^\beta 
	{\partial f\over \partial w^i} = X_{\tilde{g}}f(x,ap) ~, \]
which proves our claim. This statement is just a precise way of expressing
conformal invariance of massless particles.

\subsection{Free, massless particles in a perturbed Friedmann universe}

We now apply this general framework to the case of a perturbed Friedmann
universe. For simplicity, we restrict our analysis to the case $K=$, i.e., 
$\Om=1$. The metric of a perturbed Friedmann universe with density
parameter $\Omega=1$ is given by $ds^2=g_{\mu\nu} dx^\mu dx^\nu$ with
\begin{equation} 
 g_{\mu\nu} = a^2(\eta_{\mu\nu} + h_{\mu\nu}) = a^2 \tilde{g}_{\mu\nu}
  ~, \end{equation}
where $(\eta_{\mu\nu}) = diag(-,+,+,+)$ is the flat Minkowski metric and
$(h_{\mu\nu})$ is a small perturbation, $|h_{\mu\nu}|\ll1$.

From Eq.~(\ref{conform}), we conclude that the Liouville equation 
in a perturbed 
Friedmann universe is equivalent to the Liouville equation in perturbed
Minkowski space,
\begin{equation} (X_{\tilde{g}}f)(x,v)=0 ~, 
	\label{LM} \end{equation}
with $v=v^\mu\tilde{e}_\mu = ap^\mu\tilde{e}_\mu$.\footnote{Note that 
also Friedmann universes with non vanishing spatial curvature, 
$K\neq 0$, are conformally flat and thus 
this procedure can also be applied for $K\neq 0$. Of course, in this
case the conformal factor $a^2$ is no longer just the scale factor but
 depends on position. A coordinate transformation which transforms the
metric of $K\neq 0$ Friedmann universes into a conformally flat form
can be found, e.g., in \cite{CDD}.}

We now want to derive a linear perturbation equation for Eq.~(\ref{LM}).
If $\bar{e}^\mu$ is a tetrad in Minkowski space, 
$\tilde{e}_\mu = \bar{e}_\mu + {1\over 2}h_\mu^\nu\bar{e}_\nu$ is a 
tetrad w.r.t the 
perturbed geometry $\tilde{g}$. For 
$(x,v^\mu\bar{e}_\mu )\in  \bar{P}_0$, thus,
$(x,v^\mu\tilde{e}_\mu)\in \tilde{P}_0$. Here $\bar{P}_0$ denotes the 
zero mass one particle phase space in Minkowski space and $\tilde{P}_0$ 
is the phase space with respect to $\tilde{g}$, perturbed Minkowski 
space. We define the 
perturbation,  $F$, of the distribution function by
\begin{equation} 
   f(x,v^\mu \tilde{e}_\mu) = \bar{f}(x,v^\mu \bar{e}_\mu) + 
	F(x,v^\mu\bar{e}_\mu)
  ~. \end{equation}
Liouville's equation for $f$ then leads to a perturbation equation 
for $F$. We choose the natural tetrad 
\[\tilde{e}_\mu=\partial_\mu -{1\over 2}h_\mu^\nu\partial_\nu\]
with the corresponding basis of 1--forms
\[\tilde{\theta}^\mu=dx^\mu +{1\over 2}h^\mu_\nu dx^\nu ~.\]
Inserting this into the first structure equation, 
$d\tilde{\theta}^\mu= -\omega^\mu_{~~\nu}\wedge dx^\nu$, one finds
\[ \omega_{\mu\nu}=-{1\over 2}(h_{\mu\lambda},_\nu - 
h_{\nu\lambda},_\mu)\theta^\lambda ~.\]
Using the background Liouville equation, namely that $\bar{f}$ is
only a function of $v=ap$, we obtain the perturbation equation
\[ (\partial_t +n^i\partial_i)F = -{v\over 2}[(\dot{h}_{i0}-h_{00},_i)n^i
   +(\dot{h}_{ij}-h_{0j},_i)n^in^j]{d\bar{f}\over dv}  ~,\]
where we have set $v_i=vn_i$, with $v^2=\sum_{i=1}^3(v_i)^2$, i.e.,
$\bf n$ gives the  momentum direction of the particle.
Let us parameterize the perturbations of the metric by
\begin{equation} \left(h_{\mu\nu}\right) = \left(\begin{array}{ll} 
		-2A & B_i \\
                B_i & 2H_L\delta_{ij}
                                +2H_{ij} \end{array}\right),
 \label{scalar} \end{equation}
with $H_i^i=0$. Inserting this above we obtain
\begin{equation}
(\partial_t +n^i\partial_i)F = -[\dot{H}_L +(A,_i +{1\over 2}\dot{B}_i)n^i +
	(\dot{H}_{ij}-{1\over 2}B_{i,j})n^in^j]v{d\bar{f}\over dv} ~.
 \label {LF} \end{equation}
From Eq.~(\ref{LF}) we see that the perturbation in the distribution 
function in each spectral band is proportional to $v{d\bar{f}\over dv}$.
This shows once more that gravity is achromatic. We thus
do not loose any information if we integrate this equation over 
photon energies. We define 
\[ m = {\pi\over \rho_ra^4}\int Fv^3dv ~.\]
$4m$ is the fractional perturbation of the brightness $\iota$,
\[ \iota = a^{-4} \int f v^3dv ~. \]
Setting $\iota({\bf n,x}) = \bar{\iota}(T({\bf n,x}))$, one obtains that 
$ \iota =(\pi/60) T^4({\bf n,x})$. Hence, $m$  
corresponds to the fractional perturbation in the temperature, 
\begin{equation} T({\bf n,x}) = \bar{T}(1+m({\bf n,x})) ~.
	\label{T} \end{equation}
Another derivation of Eq.~(\ref{T}) is given in \cite{d94}. According
to Eq.~(\ref{LF}),  the $v$ dependence of  $F$ is of the form 
$v{d\bar{f}\over dv}$. Using now 
\begin{equation}   
{4\pi}\int {d\bar{f}\over dv}v^4dv = -4\int \bar{f}v^3dvd\Omega 
	=-4\rho_ra^4    ~, \label{rel}	\end{equation}
 we find
\[ F(x^\mu,n^i,v)=-m(x^\mu,n^i)v{d\bar{f}\over dv} ~. \] 
This shows that $m$ is indeed the quantity which is measured in a
CMB anisotropy experiment, where the spectral information is used
to verify that the spectrum of perturbations is the derivative of a
blackbody spectrum. Of course, in a real experiment located at a fixed
position in the Universe, the monopole and 
dipole contributions to $m$ cannot be measured. They cannot be 
distinguished from a background component and from a dipole due to our
peculiar motion w.r.t. the CMB radiation.

Multiplying Eq.~(\ref{LF}) with $v^3$ and integrating over $v$,  we obtain
the equation of motion for $m$
\begin{equation} 
	\partial_tm+n^i\partial_im= \dot{H}_L +(A,_i +{1\over 2}\dot{B}_i)n^i +
	(\dot{H}_{ij}-{1\over 2}B_i,_j)n^in^j ~.
\label{Lm}
\end{equation}

It is well known that the equation of motion for photons only couples to
the Weyl part of the curvature (null geodesics are conformally invariant).
However, the r.h.s. of Eq.~(\ref{Lm}) is given by first derivatives of 
the metric only
which could at best represent integrals of the Weyl tensor. To obtain 
a local, non integral equation, we thus rewrite Eq.~(\ref{Lm}) in terms of
$\nabla^2 m$. It turns out, that the most suitable variable is however not $\nabla^2 m$
but $\chi$, which is defined by
\[ \chi \equiv \nabla^2 m - (\nabla^2 H_L-{1\over 2}H,_{ij}^{ij}) - 
	{1\over 2}(\nabla^2 B_i-3\partial^j\sigma_{ij})n^i  ~, \]
\[ \mbox{where }~~ \sigma_{ij}\equiv -{1\over 2}(B_i,_j+B_j,_i) +
	{1\over 3}\delta_{ij}B_l^{,l} +\dot{H}_{ij}.  \]
Note that $\chi$ and $\nabla^2 m$ only differ by the monopole contribution,
$\nabla^2 H_L-(1/2)H^{ij},_{ij}$~, and the dipole term, 
$(1/2)(\nabla^2 B_i -3\partial^j\sigma_{ij})n^i$. The higher multipoles of
$\chi$ and $\nabla^2 m$ agree.
An observer at fixed position and time cannot distinguish a monopole 
contribution from an isotropic background
 and a dipole contribution from a peculiar motion.
Only the higher multipoles, $l\ge 2$ contain information about
temperature anisotropies. For a fixed observer therefore, we can 
identify $\nabla^{-2}\chi$ with $\delta T/T$.

In terms of  metric perturbations, the electric and magnetic part of the 
Weyl tensor are given by ({\sl see}, e.g. \cite{Ma,d94})
\begin{eqnarray}
 {\cal E}_{ij} &=&  {1\over 2}[\triangle_{ij}(A-H_L) -\dot{\sigma}_{ij}
		-\nabla^2 H_{ij}-{2\over 3}H_{lm}^{,lm}\delta_{ij}
	+ H_{il}^{,l},_j + H_{jl}^{,l},_i] \label{E} \\
 {\cal B}_{ij} &=& -{1\over 2}(\epsilon_{ilm}\sigma_{jm},_l + \epsilon_{jlm}\sigma_{im},_l ) ~,
  \label{B}  \end{eqnarray}
\[ \mbox { with }~~ \triangle_{ij} =\partial_i\partial_j -(1/3)\delta_{ij}\nabla^2 ~.\]

Explicitly working out $(\partial_t+n^i\partial_i)\chi$ using Eq.~(\ref{Lm}),
 yields after some algebra the equation of motion for $\chi$:
\begin{equation}
 (\partial_t +n^i\partial_i)\chi = 3n^i\partial^j{\cal E}_{ij} +n^kn^j\epsilon_{kli}
	\partial_l{\cal B}_{ij} \equiv {\cal S}(t,\mbox{\boldmath $x$},\mbox{\boldmath $n$})~ ,
  \label{Lchi}  \end{equation}
where $\epsilon_{kli}$ is the totally antisymmetric tensor in three 
dimensions with $\epsilon_{123}=1$.
The spatial indices in this equation are raised and lowered with 
$\delta_{ij}$ and thus  index positions are irrelevant. Double indices are
summed over irrespective of their positions.

Eq. (\ref{Lchi}) is the main result of this paper. We now discuss it,
rewrite it in integral form and specify initial conditions for adiabatic
scalar perturbations with or without seeds.

In Eq.~(\ref{Lchi}) the contribution from the electric part of the
Weyl tensor is a divergence, and therefore does not contain tensor perturbations.  On the other hand, 
scalar perturbations do not induce a magnetic gravitational field. The
second contribution to the source term in Eq.~(\ref{Lchi}) thus represents a 
combination of vector and tensor perturbations. If vector perturbations
are negligible (like, e.g., in models where initial fluctuations are
generated during an epoch of inflation), the two terms on the r.h.s of
Eq.~(\ref{Lchi}) 
yield thus a split into scalar and tensor  perturbations which is local.

Since the Weyl tensor of Friedmann Lema\^{\i}tre universes vanishes, the
r.h.s. of Eq.~(\ref{Lchi}) is manifestly gauge invariant (this is the so 
called Stewart--Walker lemma \cite{SW}). Hence also the variable
$\chi$ is gauge invariant. 
Another proof of the gauge 
invariance of $\chi$, discussing the behavior of $F$ under infinitesimal
coordinate transformations is presented in \cite{d94}.

The general solution of Eq.~(\ref{Lchi}) is given by

\begin{equation} 
	\chi(t,\mbox{\boldmath $x$},\mbox{\boldmath $n$}) =
	\int_{t_i}^t {\cal S}(t',\mbox{\boldmath $x$}+(t'-t)\mbox{\boldmath $n$}, \mbox{\boldmath $n$})dt'
	~ + ~ \chi(t_i,\mbox{\boldmath $x$}+(t_i-t)\mbox{\boldmath $n$}, \mbox{\boldmath $n$}) ~,
\label{chi} \end{equation}
where ${\cal S}$ is the source term on the r.h.s. of Eq.~(\ref{Lchi}).

In Appendix~A we derive the relations between the geometric source
term ${\cal S}$ and the energy momentum tensor in a
perturbed Friedmann universe. 

\section{The collision term}
In order for Eq.~(\ref{chi}) to provide a useful solution, we need to
determine the correct initial conditions, $\chi(t_{dec})$, at the 
moment of
decoupling of matter and radiation. Before recombination, photons,
electrons and baryons form a tightly coupled plasma, and thus $\chi$
can not develop higher moments in $\bm n$. The main collision
process is non--relativistic Compton scattering of electrons and
photons. The only non vanishing moments in the distribution function
before decoupling are the zeroth, i.e., the energy density, and the first, the
energy flow. We therefore set
\be \chi(t_{dec})=\nabla^2\left({1\over4}D^{(r)}_g(t_{dec}) 
	-\bm{n}\cdot\bm{V}^{(r)}(t_{dec})\right)  ~, \label{initial}
\ee
where
\bea
 D^{(r)}_g(t_{dec}) &=&
	\nabla^{-2}\left({1\over\pi}\int\chi(t_{dec})d\Om \right)\\
	&=& {\de\rho^{(r)}\over \rho} -4H_L +2\nabla^{-2}(H_{ij}^{|ij})
  ~~\mbox{ and}  \nonumber\\
 V^{j(r)}(t_{dec}) &=& 
	-\nabla^{-2}\left({3\over4\pi}\int\chi(t_{dec})n^jd\Om
	\right) \label{Veq}\\ 
	&=& -T^{(r)j}_0/({4\over 3}\rho^{(r)}) + B^i - {3\over 2}
	\nabla^{-2}(\dd^i\si_{ij}) ~. \nonumber\eea
$D_g^{(r)}$ and $\bm{V}^{(r)}$ are gauge invariant density and velocity
perturbation variables \cite{KS,d94}.

In the tight coupling or fluid limit, the initial conditions can also
be obtained from the collision term. Setting ${\cal M}\equiv\nabla^{-2}\chi$
one finds the following expression for the collision integral
\cite{d94},
\be C[{\cal M}] = a\si_Tn_e[{1 \over 4}D_g^{(r)} -{\cal M} + {\bm
	n\cd V}^{(b)} + {1\over 2}n_in_jM^{ij}] ~. \label{col} \ee
The last term is due to the anisotropy of the cross section for 
non--relativistic Compton scattering, with 
\[ M^{ij} = {3\over 8\pi}\int(n^in^j-{1\over 3}\de^{ij}){\cal M}d\Om
~.\]
$\cal M$ is a gauge invariant perturbation variable for the
distribution function of photons. $\bm{V}^{(b)}$ denotes the baryon
velocity field,  $\si_T$ and $n_e$ are
 the Thomson cross section and the free electron density respectively.
 To make contact with other literature, 
we note that ${\cal M} = \Theta +\Phi$, where $\Theta$ is the 
perturbation variable describing the CMB anisotropies defined in 
\cite{HuSu1} and $\Phi$ denotes a Bardeen potential ({\sl see Section~4}).
 Since $\cal M$ and $\Theta$ differ only by a monopole term, 
they give rise to the same spectrum of temperature anisotropies for $\ell\ge 1$.
$\cal M$ satisfies the
Boltzmann equation
\be (\dd_t +n^i\dd_i){\cal M} =\nabla^{-2}{\cal S} +C[{\cal M}] ~,
	\label{Bol}\ee
where $\cal S$ is the gravitational source term given in
Eq.~(\ref{Lchi}). 
In the tight coupling limit, $t_T\equiv (a\si_Tn_e)^{-1}\ll
t$, we may, to lowest order in $(t_T/t)$, just set the square bracket
on the right hand side of Eq.~(\ref{col}) equal to zero. Together with
Eq.~(\ref{Veq}) this yields
\[ \bm{V}^{(b)} =\bm{V}^{(r)} ~.\]
Neglecting gravitational effects, the right hand side of  Boltzmann's
equation then leads to
\be \dot{D}^{(r)}_g = {4\over 3}\bm{\nabla\cd V}^{(b)} = 
	{4\over 3}D_g^{(b)}~,\label{Dg} \ee 
where the last equal sign is due to baryon number conservation. In
other words, photons and baryons are adiabatically coupled. Expanding
Eq.~(\ref{Bol}) one order higher in $t_T$, one obtains Silk 
damping \cite{Silk},
the damping of radiation perturbations due to imperfect coupling.

Let us estimate this damping by neglecting gravitational effects and
the time dependence of the coefficients in the Boltzmann equation
(\ref{Bol}) since we are interested in time scales $t_T\ll t$. We can
then look for solutions of the form
\[ V^{(b)} \propto {\cal M} \propto \exp(i(\bm{kx}-\om t))  ~.\]
We also neglect the angular dependence of the collision term. Solving
Eq.~(\ref{Bol}) for $\cal M$, we then find
\be {\cal M} = {(1/4)D_g^{(r)} +i\bm{k\cd n}V^{(b)}\over
	1-it_T(\om-\bm{k\cd n})}  ~.  \label{mmm}\ee
The collisions also induce a drag force in the equation of motion of
the baryons which is given by
\[ F_i={a\si_Tn_e\rho_r\over \pi}\int C[{\cal M}]n_id\Om = 
	{4\rho_r\over 3t_T}(\bm{V}^{(r)}-i\bm{k}V^{(b)}) ~.\]
With this force, the baryon equation of motion becomes
\[ \bm{k}\om V^{(b)} +i(\dot{a}/a)\bm{k}V^{(b)} =i\bm{k}\Psi -\bm{F}/\rho_b ~.\]
To lowest order in $t_T/t$ and $kt_T$, this leads to the following
correction  to the adiabatic condition $\bm{V}^{(b)}=\bm{V}^{(r)}$:
\be	t_T\om\bm{k} V^{(b)}  =   {4\rho_r\over 3\rho_b}
	(i\bm{k}V^{(b)}-\bm{V}^{(r)}) ~, \label{vvv}
\ee
From Eq.~(\ref{Dg}) we obtain the relation $\bm{k\cd V}^{(r)} = -(3/4)\om
D^{(r)}_g$ to lowest order. Using this approximation, we find,
after multiplying  Eq.~(\ref{vvv})  with $\bm{k}$,
\be V^{(b)}= {(3/4)\om\over t_Tk^2\om R-ik^2}D_g^{(r)} ~, \label{V}\ee
with $R=3\rho_b/\rho_r$. The densities $\rho_b$ and $\rho_r$ denote
the baryon and radiation densities respectively. Inserting this result
in Eq.~(\ref{mmm}) leads to
\be {\cal M} = {1 +{3\mu\om/k\over 1-it_T\om R}\over
	1-it_T(\om-k\mu)}D_g^{(r)}/4 \label{MMM} ~,\ee
where we have set $\mu=\bm{k\cd n}/k$. From this result, which is
valid on time scales shorter than the expansion time (length scales
smaller than the horizon), we can derive a dispersion relation
$\om(k)$. In lowest order $\om t_T$ we obtain
\bea
 \om=\om_0 -i\ga &\mbox{with}&\\
\om_0 ={k\over \sqrt{3(1+R)}} &\mbox{and}& \ga=k^2t_T{R^2+{4\over
	5}(R+1)\over 6(R+1)^2} ~. \label{disp}\eea
At recombination $R\sim 0.1$ so that $\ga \sim 2k^2t_T/15$. 

We have thus found that, due to
diffusion damping, the photon perturbations thus undergo an
exponential decay which can be approximated by
\be |{\cal M}| \propto \exp(-2k^2t_Tt/15) ~, \mbox{ on scales }~ 
	t\gg 1/k\gg t_T ~.\label{silk} \ee

In general, the temporal evolution of radiation perturbations can be split 
into three regimes:
Before recombination, $t\ll t_{dec}$ the evolution of photons can be
determined in the fluid limit. After recombination, the free Liouville
equation is valid. Only during recombination the full Boltzmann
equation has to be considered, but also there collisional damping can be
reasonable well approximated by an exponential damping envelope 
\cite{HuSu}, which is a somewhat sophisticated version of (\ref{silk}). 

\section{Example: Adiabatic scalar perturbations} 
We now want to discuss Eq.~(\ref{Lchi}) with initial conditions given
by Eq.~(\ref{initial}) in some examples.

Perturbations are called 'scalar' if all 3 dimensional tensors
(tensors w.r.t their spatial components on  hyper-surfaces of
constant time) can be obtained as derivatives of scalar potentials. 

Scalar perturbations of the geometry can be described by two gauge
invariant variables, the Bardeen potentials \cite{Ba} $\Phi$ and
$\Psi$. The variable $\Psi$ is the relativistic analog of the
Newtonian potential. In the Newtonian limit, $-\Phi=\Psi$= the Newtonian
gravitational potential. In the relativistic situation, $\Phi$ is better
interpreted as the perturbation in the  scalar curvature on the
hyper-surfaces of constant time \cite{dS}. In terms of the Bardeen 
potentials, the electric and magnetic
components of the Weyl tensor are given by \cite{Ma}
\be {\cal E}_{ij} = {1\over 2}\lap_{ij}(\Phi-\Psi)
	~~,~~~ {\cal B}_{ij}=0 ~, \label{Escal}\ee
where $\lap_{ij}$ denotes the traceless part of the second derivative,
$\lap_{ij}= \dd_i\dd_j-{1\over 3}\de_{ij}\nabla^2$. The Liouville
equation, (\ref{Lchi}) then reduces to
\be (\dd_t +n^i\dd_i){\cal M} = n^i\dd_i(\Phi-\Psi) ~. \ee
With the initial conditions given in Eq.~(\ref{initial}) we find the
solution
\be {\de T\over T}(t_0,\bm{x}_0,\bm{n})= {\cal M}(t_0,\bm{x}_0,\bm{n})
	= [{1\over 4}D_g^{(r)} +n^i\dd_iV^{(b)}
	+\Psi-\Phi](t_{dec},\bm{x}_{dec}) -
	\int_{t_{dec}}^{t_0}(\dot{\Phi}-\dot{\Psi})(t,\bm{x}(t))dt ~,
\label{sols} \ee
where $\bm{x}_{dec}=\bm{x}_0-(t_0-t_{dec})\bm{n}$ and correspondingly
$\bm{x}(t)=\bm{x}_0-(t_0-t)\bm{n}$.

We now want to replace the fluid variables, $D_g^{(r)}$ and $V^{(b)}$, 
wherever possible, by
perturbations in the geometry.  To this goal, let us first consider 
the general situation, when one part of the geometry perturbation is 
due to perturbations in the cosmic matter components
 and another part is due to some type of seeds, which do not
contribute to the background energy and pressure. The Bardeen
potentials can then be split into contributions from matter and seeds:
\be \Phi = \Phi_m +\Phi_s ~~, \Psi = \Psi_m +\Psi_s ~.\ee
To proceed further, we must assume a relation between the
perturbations in the total energy density and energy flow, $D_g$ and
$V$, and the corresponding perturbations in the photon component. The
most natural assumption here is that perturbations are adiabatic,
i.e., that 
\[ D_g^{(r)}/(1+w_r) = D_g/(1+w) ~ \mbox{ and }~~~V^{(b)}=V^{(r)} = V ~,\]
where $w\equiv p/\rho$ denotes the enthalpy, i.e. $w_r=1/3$. 
For $w_r \neq w$ this condition can only be maintained on super--horizon 
scales or for tightly coupled fluids. For decoupled fluid
components, the different equations of state lead to a violation of
this initial condition on sub--horizon scales.

In order to use the perturbed Einstein equations to replace $D_g$ and
$V$ by geometric perturbations we define yet another density
perturbation variable, 
\bean 
	D &\equiv& D_g +3(1+w){\dot{a}\over a}V -3(1+w)\Phi ~~~\mbox{and}\\
	D^{(r)} &\equiv& D_g^{(r)} +4{\dot{a}\over a}V^{(r)} -4\Phi ~.
\eean
The matter perturbations $D$ and $V$ determine the matter part of the
Bardeen potentials via the perturbed Einstein equations ({\sl see},
e.g. \cite{d94}). The following relation between $\Phi_m$
and $D$ can also be obtained using Eqs.~(\ref{Escal}) and (A16) in the
absence of seeds. 
\[ D =- {2\over 3}\left({\dot{a}\over a}\right)^{-2}\nabla^2\Phi_m
\sim (kt)^2\Phi_m  ~~\mbox{ and}\]
\[ {\dot{a}\over a}\Psi_m -\dot{\Phi}_m= {3\over 2}\left({\dot{a}\over
	a}\right)^2(1+w)V ~.  \]
The term $D$ rsp. $D^{(r)}$, is much smaller than the
Bardeen potentials on super--horizon scales and it starts to dominate
on sub--horizon scales, $kt\gg 1$. For this term therefore, the
adiabatic relation is not useful and we should not replace $D^{(r)}$
by ${4\over 3(1+w)}D$. The same holds for $\dd_iV^{(b)}$ which is of
the order of $kt\Phi_m$. However, $(\dot{a}/a)V^{(r)}$ is of the same
order of magnitude as the Bardeen potentials and thus mainly relevant
on super horizon scales. There the adiabatic condition makes sense and
we may replace $(\dot{a}/a)V$ by its expression in terms geometric 
perturbations. Keeping only $D^{(r)}$ and $\dd_iV^{(b)}$
 in terms of photon fluid variables, Eq.~(\ref{sols}) becomes
\bea {\de T\over T}(\bm{x}_0,t_0,\bm{n}) &=& [ \Psi_s + {1+3w\over 3+3w}\Psi_m + 
	{2\over 3(1+w)}\left({\dot{a}\over a}\right)^{-1}\dot{\Phi}_m
	 + {1\over 4}D^{(r)} + n^i\dd_iV^{(b)}](\bm{x}_{dec},t_{dec}) \nonumber\\
  && -\int_{t_{dec}}^{t_0}(\dot{\Phi}-\dot{\Psi})(\bm{x}(t),t) ~.
 \label{solgen}\eea

This is the most general result for adiabatic scalar perturbations in
the photon temperature. It contains geometric perturbations, acoustic
oscillations prior to recombination and the Doppler term.  Silk
damping, which is relevant on very small angular scales (see the
contribution by \cite{La}) is neglected, i.e., we assume
'instantaneous recombination'. Eq.~(\ref{solgen}) is valid for all types of
matter models, with or without cosmological constant and/or spatial
curvature (we just assumed that the latter is negligible at the last
scattering surface, which is clearly required by observational
constraints). The first two terms in the square bracket are usually 
called the ordinary Sachs--Wolfe contribution. The integral is the
'integrated Sachs--Wolfe effect'. The third and fourth
term in the square bracket describe the acoustic Doppler oscillations
respectively. On super horizon scales, $kt\ll 1$, they can be neglected. 

To make contact with the formula usually found in textbooks, we
finally constrain ourselves to a universe dominated by
cold dark matter (CDM), i.e., $w=0$ without any seed perturbations. In
this case $\Psi_s=\Phi_s=0$ and it is easy to show that
$\Psi=-\Phi$ and that $\dot{\Phi}=\dot{\Psi}=0$ (see, e.g.,
\cite{d94}).  Our results then
simplifies on super--horizon scales, $kt\ll 1$, to the well--known
relation of Sachs and Wolfe \cite{SaW}
\be \left({\de T\over T}\right)_{SW} = 
	{1\over 3}\Psi(\bm{x}_0-t_0\bm{n},t_{dec}) ~.\ee

\section{Conclusions}
We have derived all  the basic ingredients  to
determine the  temperature fluctuations in the CMB. Since the fluctuations are
so small, they can be calculated fully within linear cosmological
perturbation theory. Note however that density perturbations along the
line of sight to the last scattering surface might be large, and thus the
Bardeen potentials inside the Sachs Wolfe integral might have to be
calculated within non--linear Newtonian gravity. But the Bardeen
potentials themselves remain small (as long as the photons never come 
close to black holes) such that Eq.~(\ref{solgen}) remains valid. In
this way, even a CDM model can lead to an integrated Sachs Wolfe
effect which then is known under the name 'Rees Sciama
effect'. Furthermore, do to ultra violet  radiation of the first objects formed
by gravitational collapse, the universe might become reionized and
electrons and radiation become coupled again. If this reionization
happens early enough ($z>30$) the subsequent collisions lead to
additional damping of anisotropies on angular scales up to about
$5^o$. However, present CMB anisotropy measurements do not support
early reionization and the Rees Sciama effect is probably very small.
Apart from these effects due to non--linearities in the matter 
distribution, which depend on the details of the structure formation 
process, CMB anisotropies can be determined within linear perturbation 
theory.

This is one of the main reason, why observations of CMB anisotropies
may provide detailed information about the cosmological parameters
(see contribution by A. Lasenby): The main physics is linear and well known
and the anisotropies can thus  be calculated within an accuracy
of 1\% or so. The detailed results do depend in several ways on the
parameters of the cosmological model which can thus be determined by
comparing calculations with observations. 

There is however one caveat:
If the perturbations are induced by seeds (e.g. topological defects),
the evolution of the seeds themselves is in general non--linear and
complicated. Therefore, much less accurate predictions have been made
so far for models where perturbations are induced by seeds 
(see, e.g., \cite{DZ,TC,DGS}). In this
case, the observation of CMB anisotropies might not help very much to
constrain cosmological parameters, but it might contain very
interesting information about the seeds, which according to present
understanding originate from very high temperatures, $T\sim
10^{16}$GeV. The CMB anisotropies might thus bury some 'fossils' of
the very early universe, of the physics at an energy scale which we can
never probe directly by accelerator experiments.

\newpage
\appendix
\setcounter{equation}{0}
\renewcommand{\theequation}{A\arabic{equation}}

\section{An equation of motion for the  Weyl 
	tensor} 
The Weyl tensor of a spacetime $({\cal M},g)$ is defined by
\begin{equation} 
	C^{\mu\nu}_{~\;\;\sigma\rho}= R^{\mu\nu}_{~\;\;\sigma\rho} 
	-2g^{[\mu}_{~\;[\sigma}R^{\nu]}_{~\; \rho]}
	+{1\over 3}Rg^{[\mu}_{~\;[\sigma}g^{\nu]}_{\;~ \rho]} ~,
	\label{weyl} \end{equation}
where $[\mu ... \nu]$ denotes anti-symmetrization in the indices $\mu$
and $\nu$.
The Weyl curvature has the same symmetries as the Riemann curvature
and it is traceless. In addition the Weyl tensor  is 
invariant under conformal transformations:
\[ C^{\mu}_{\;\;\nu\sigma\rho}(g)= C^{\mu}_{\;\;\nu\sigma\rho}(a^2g) \]
(Careful: This equation only holds for the given index position.) 
In four dimensional spacetime, the Bianchi identities together with
 Einstein's equations yield equations of motion for the Weyl curvature.
In four dimensions, the Bianchi identities,
\[ R_{\mu\nu[\sigma\rho;\lambda]} = 0 \]
are equivalent to  \cite{CDD}
\begin{equation} C^{\alpha\beta\gamma\delta};_\delta = R^{\gamma [\alpha;\beta ]} -{1\over 6}
 g^{\gamma [ \alpha}R^{;\beta ]}  ~. \label{bianchi}\end{equation}
This together with Einstein's equations yields
\begin{equation} C^{\alpha\beta\gamma\delta};_\delta = 8\pi G(T^{\gamma[\alpha;\beta]} -{1\over 3}
 g^{\gamma[\alpha}T^{;\beta]})  ~, \label{einstein}\end{equation}
where $T_{\mu\nu}$ is the energy momentum tensor, $T=T^\lambda_\lambda$.

Let us now choose some time-like 
unit vector field $u$, $u^2=-1$. We then 
can  decompose any  tensor field into longitudinal and transverse 
components with respect to $u$. We define
\[ h^\mu_{~\nu} \equiv g^\mu_{~\nu} +u^\mu u_\nu ~, \]
 the projection onto the subspace of tangent space normal to $u$.
The decomposition of the Weyl tensor yields its electric and magnetic 
contributions:
\begin{eqnarray} {\cal E}_{\mu\nu} &=& 
	C_{\mu\lambda\nu\sigma}u^\lambda u^\sigma \\
 {\cal B}_{\mu\nu} &=& {1\over 2}C_{\mu\lambda\gamma\delta}u^\lambda\
	\eta^{\gamma\delta}_{\;\;\nu\sigma} u^\sigma ~;\end{eqnarray}
where $\eta^{\alpha\beta\gamma\delta}$ denotes the totally
antisymmetric 4 tensor with $\eta_{0123}=\sqrt{-g}$.
Due to  symmetry properties and the tracelessness of the Weyl 
curvature, $\cal E$ and $\cal B$ are symmetric and traceless, and they fully 
determine the Weyl curvature. One easily checks that 
${\cal E}_{\mu\nu}$ and
${\cal B}_{\mu\nu}$ are also conformally invariant.
We now want to perform the corresponding decomposition for the energy 
momentum  tensor of some arbitrary type of seed, $T^{S}_{\mu\nu}$.
We define
\begin{eqnarray}
\rho_S &\equiv& T^{(S)}_{\mu\nu}u^\mu u^\nu     \\
p_S &\equiv& {1\over 3}T^{(S)}_{\mu\nu}h^{\mu\nu}     \\
q_\mu &\equiv& -h_\mu^{~\nu}T^{(S)}_{\nu\alpha}u^\alpha ~~~ 
	~~~  q_i=-{1\over a}T^{(S)}_{0i} \\
\tau_{\mu\nu} &\equiv& h_\mu^{~\alpha}h_\nu^{~\beta}T^{(S)}_{\alpha\beta}-
	h_{\mu\nu}p_S ~. \end{eqnarray}
We then can write
\begin{equation} T_{\mu\nu}^{(S)}=\rho_Su_\mu u_\nu +p_Sh_{\mu\nu} +q_\mu u_\nu
	+u_\mu q_\nu +\tau_{\mu\nu} ~. \label{Tsplit} \end{equation}
This is the most general decomposition of a symmetric second rank tensor.
It is usually interpreted as the energy momentum tensor of an imperfect
fluid. In the frame of an observer moving with four velocity $u$,
$\rho_S$ is the energy density, $p_S$ is the 
isotropic pressure, $q$ is
the energy flux, $u\cdot q=0$, and $\tau$ is the tensor of anisotropic
stresses, $\tau_{\mu\nu}h^{\mu\nu}=\tau_{\mu\nu}u^\mu=0$. 

We now want to focus on
a perturbed Friedmann universe. We therefore  consider
a four velocity field $u$ which deviates only in first order from the
Hubble flow: $u=(1/a)\partial_0 +$ first order. Friedmann universes are conformally
flat, and we require the seed to represent a small perturbation on a universe
dominated by radiation and cold dark matter (CDM). The seed energy momentum
tensor  and the Weyl tensor are of thus of first order,
and (up to first order) their
decomposition does not depend on the choice of the first order 
contribution to $u$, they are gauge--invariant. But the decomposition
of the dark matter depends on this choice. Cold dark matter is
a pressure-less perfect fluid We can thus choose $u$ to denote the energy
flux of the dark matter, $T^\mu_\nu u^\nu = -\rho_{C} u^\mu$. Then the
energy momentum tensor of the dark matter has the simple decomposition
\begin{equation} T^{(C)}_{\mu\nu} = \rho_{C}u_\mu u_\nu  \label{DM} ~. \end{equation}
With this choice, the Einstein equations Eq.~(\ref{einstein}) linearized
about an $\Omega=1$ Friedmann background 
yield the following 'Maxwell equations' for $E$ and $B$ \cite{El}:\\
{\em i) Constraint equations}
\begin{eqnarray}  
\partial^i{\cal B}_{ij} &=& 4\pi G \eta_{j\beta\mu\nu}u^\beta q^{[\mu;\nu]}  
	\label{diB}\\
\partial^i{\cal E}_{ij} &=& 8\pi G( {1\over 3}a^2\rho_{C}D,_j +{1\over 3}a^2\rho_S,_j 
	-{1\over 2}\partial^i\tau_{ij} -{\dot{a}\over a^2}q_j) ~.
	\label{diE}\end{eqnarray}
{\em ii) Evolution equations}
\begin{eqnarray}  
a\dot{\cal B}_{ij} +\dot{a}{\cal B}_{ij} -a^2h_{(i}^{~~\alpha}
	\eta_{j)\beta\gamma\delta}u^\beta {\cal E}_{\alpha}^{~~\gamma;\delta}  
    &=&- 4\pi Ga^2h_{\alpha(i}\eta_{j)\beta\mu\nu}u^\beta\tau^{\alpha\mu;\nu}
	  \label{dtB} \\
\dot{\cal E}_{ij} +{\dot{a}\over a}{\cal E}_{ij} +ah_{(i}^{~~\alpha}
	\eta_{j)\beta\gamma\delta}u^\beta{\cal B}_{\alpha}^{~~\gamma;\delta}  
	&=& -4\pi G(aq_{ij}  -{\dot{a}\over a}\tau_{ij} +\dot{\tau}_{ij} 
	+a\rho_{C}u_{ij}) ,
	\label{dtE}\end{eqnarray}
where $(i ...j)$ denotes symmetrization in the indices $i$ and $j$. 
The symmetric traceless tensor fields $q_{\mu\nu}$ and $u_{\mu\nu}$ 
are defined by
\begin{eqnarray*}
	q_{\mu\nu}&=& q_{(\mu;\nu)}-{1\over 3}h_{\mu\nu}q^\lambda_{~;\lambda}\\
  u_{\mu\nu}&=& u_{(\mu;\nu)}-{1\over 3}h_{\mu\nu}u^\lambda_{~;\lambda} ~.
\end{eqnarray*}
In Eqs.~(\ref{dtB}) and (\ref{dtE})
we have also used that for the dark matter perturbations only scalar
perturbations are relevant, vector perturbations decay quickly. Therefore
$u$ is a gradient field, $u_i =U_{; i}$ for some suitably chosen 
function $U$. Hence the vorticity of the vector field $u$ vanishes,
$u_{[\mu;\nu]}=0$. With
\[ \eta_{0ijk} =a^4\epsilon_{ijk} ~~,~~~ \rho_S=a^{-2}T^{S}_{00} ~~
	\mbox{ and }~~ q_i = -a^{-1}T_{0i}^{S}  ~,\]
we obtain from Eq.~(\ref{diE})
\begin{equation}
\partial^i{\cal E}_{ij} = 8\pi G( {1\over 3}\rho_{C}a^2D,_j +{1\over 3}
	T_{00}^{S},_j
	-{1\over 2}\partial_i\tau_{ij} +{\dot{a}\over a}T_{0j}^{S}) ~.
 \label{dEj}\end{equation}
In Eq.~(\ref{dEj}) and the following
equations summation over double indices is understood, irrespective of 
their position.

To obtain the equation of motion for the magnetic part of the
Weyl curvature we take the time derivative of Eq.~(\ref{dtB}), using
 $u=(1/a)\partial_0 +1.$order and $\eta_{0ijk}=a^4\epsilon_{ijk}$. This leads to
\begin{equation}
 (a{\cal B}_{ij})^{\cdot\cdot} = -a(\epsilon_{lm(i}[\dot{\cal E}_{j)l}
	+{\dot{a}\over a}{\cal E}_{j)l}],_m -4\pi G
	\epsilon_{lm(i}[\dot{\tau}_{j)l},_m + 
	{\dot{a}\over a}\tau_{j)l},_m])  ~,
\label{ddB} \end{equation}
where we have again used that $u$ is a gradient field and thus terms
like $\epsilon_{ijk}u_{lj},_k$ vanish.  We now insert Eq.~(\ref{dtE}) into the
first square bracket above and  replace
product expressions of the form $\epsilon_{ijk}\epsilon_{ilm}$
and $\epsilon_{ijk}\epsilon_{lmn}$ with double and triple Kronecker
deltas. Finally we replace divergences of $B$ with the help of 
Eq.~(\ref{diB}). After some algebra, one obtains
\[ \epsilon_{lm(i}[\dot{\cal E}_{j)l} +{\dot{a} \over a}{\cal E}_{j)l}],_m 
	= -\nabla^2 {\cal B}_{ij}
	-4\pi G\epsilon_{lm(i}[2aq_l,_{mj)} + \dot{\tau}_{j)l},_m -
	{\dot{a}\over a^2}\tau_{j)l},_m] ~.\]
Inserting this into Eq.~(\ref{ddB}) and using energy momentum
conservation of the seed, 
we finally find the equation of motion for $\cal B$:
\begin{equation}
 a^{-1}(a{\cal B})^{\cdot\cdot}_{ij}   -\nabla^2 {\cal B}_{ij}= 8\pi G{\cal S}^{(B)}_{ij} ~,
\label{AB}\end{equation}
with
\begin{equation}
{\cal S}^{(B)}_{ij} = \epsilon_{lm(i}[-T^{S}_{0l},_{j)m} +
	\dot{\tau}_{j)l},_m]~.
\end{equation}
Eq.~(\ref{AB}) is the linearized wave equation for the magnetic part of 
the Weyl tensor in an expanding universe. A similar equation can also be 
derived for $\cal E$. 
 
 Since dark matter just induces
scalar perturbations and  ${\cal B}_{ij}$ is sourced by vector and tensor 
perturbations only, it is  independent of the dark matter fluctuations.
Equations Eqs.~(\ref{dEj}) and (\ref{AB}) connect the source terms in the 
Liouville equation of section~2, $\partial^i{\cal E}_{ij}$ and 
${\cal B}_{ij}$ to the perturbations of the energy momentum tensor.  
\newpage


\begin{thebibliography}{99}
\bibitem{Sm}G.F. Smoot   et al., {\em Astrophys. J.} {\bf 396}, L1 (1992);
  E.L.Wright, et al. {\it Astrophys. J.} {\bf 396}, L13 (1992).
\bibitem{HZ} E. Harrison, {\em Phys. Rev.} {\bf D1} 2726 (1970);\\
	 Ya. B. Zel'dovich, {\em Mont. Not. R. Astr. Soc.} 
        {\bf 160}, P1 (1972).
\bibitem{Slava}V.F. Mukhanov, R.H. Brandenberger and H.A. Feldmann, 
	{\em Phys. Rep.} {\bf 215}, 203   (1991).
\bibitem{Kibble}T. Kibble, {\em Phys. Rep.} {\bf 67}, 183 (1980).
\bibitem{mor} G. Smoot and D. Scott in: L. Montanet et al.,
	 {\em Phys. Rev.} {\bf D50},
	1173 (1994), 1996 upgrade, available at URL: http://pdg.lbl.gov; or
	astro-ph/9603157.
\bibitem{La} A. Lasenby, these Proceeedings.
\bibitem{Ste} J.M. Stewart, {\em Non-Equilibrium Relativistic  Kinetic 
	Theory}, Springer Lecture Notes in Physics, 
	Vol. 10, ed. J. Ehlers, K. Hepp and H.A. Wiedenm\"uller
	(1971). 
\bibitem{CDD}Y. Choquet--Bruhat, C. De Witt--Morette and
	M. Dillard--Bleick, {\em Analysis, Manifolds and Physics},
	North--Holland (Amsterdam, 1982).
\bibitem{KS}H. Kodama and M. Sasaki, {\em Theor. Phys. Suppl.} {\bf
	78} (1980).
\bibitem{d94}R. Durrer, {\em Fund. of Cosmic Physics} {\bf 15}, 209 (1994).
\bibitem{Ma}J.C.R. Magueijo, {\em Phys. Rev.} {\bf D46}, 3360 (1992).
\bibitem{SW}J.M. Stewart and M. Walker, {\em Proc. R. Soc. London}
{\bf A341}, 49 (1974).
\bibitem{HuSu1}W. Hu and N. Sugiyama, {\em Phys. Rev.} {\bf D51}, 2599
	(1995).
\bibitem{Silk}J. Silk, {\em Astrophys. J.} {\bf 151}, 459 (1968).
\bibitem{HuSu}W. Hu and N. Sugiyama, ``Small scale cosmological 
	perturbations: an analytic approach'', astro-ph/9510117 (1995).
\bibitem{Ba}J. Bardeen, {\em Phys. Rev.} {\bf D22}, 1882 (1980).
\bibitem{dS}R. Durrer and N. Straumann, {\em Helv. Phys. Acta} {\bf
	61}, 1027 (1988).
\bibitem{SaW}R.K. Sachs and A.M. Wolfe, {\em Astrophys. J.} {\bf 147},
	73 (1967).
\bibitem{DZ}R. Durrer and Z.H. Zhou, {\em Phys. Rev.} {\bf D53}, 5394
	(1996).
\bibitem{TC}R.G. Crittenden and N. Turok {\em Phys. Rev. Lett.} {\bf
	75}, 2642 (1995).
\bibitem{DGS}R. Durrer, A. Gangui and M. Sakellariadou, {\em
Phys. Rev. Lett.} {\bf 76}, 579 (1996).
\bibitem{El}G. Ellis, F.R.S., in: {\em Varenna Summer School on General
	Relativity and Cosmolocy} XLVII Corso, Academic Press 
	(New York, 1971).
\end{thebibliography}
\end{document}